\documentstyle[12pt]{article}
\begin{document}
\begin{center}
{\Large {\bf MESONIC AND BINDING CONTRIBUTIONS TO THE EMC EFFECT IN A
RELATIVISTIC MANY BODY APPROACH}}
\end{center}

\vspace{0.4cm}
\begin{center}
{\Large {E. Marco$^1$, E. Oset$^1$ and P. Fern\'andez de C\'ordoba$^2$}}
\end{center}

\vspace{0.7cm}

$^1$ {\small{\it Departamento de F\'{\i}sica Te\'orica and IFIC, 
Centro Mixto Universidad de Valencia-CSIC, 
46100 Burjassot (Valencia), Spain}}

$^2$ {\small{\it Departamento de Matem\'atica Aplicada, Universidad Polit\'ecnica
de Valencia}}

\begin{abstract}
{\small}
We revise the conventional nuclear effects of Fermi motion, binding and pionic
effects in deep inelastic lepton scattering using a relativistic formalism
for an interacting Fermi sea and the local density approximation to translate
results from nuclear matter to finite nuclei. In addition we also consider
effects from $\rho$-meson renormalization in the nucleus.
The use of nucleon Green's 
functions in terms of their spectral functions offers a precise way to account
for Fermi motion and binding. On the other hand the use of many body Feynman
diagrams in a relativistic framework allows one to avoid using prescriptions
given in the past
to introduce relativistic corrections in a non relativistic
formalism.

We show that with realistic nucleon spectral functions and meson nucleus 
selfenergies one can get a reasonable description of the EMC effect for
$x > 0.15$, outside the shadowing region. 
\end{abstract}

\newpage

\section{Introduction}
The EMC effect \cite{1} is probably one of the topics in the interplay
of nuclear and particle physics which has attracted more attention. Pioneering
work on Fermi motion and pionic effects \cite{2,3,3A} was followed by many
different ideas like binding of the nucleons in the nucleus, multiquark
cluster effects or $Q^2$ rescaling (see refs. \cite{4,5,6} for reviews on 
the topic). Here we shall pay attention only to conventional nuclear degrees
of freedom, mesons ($\pi$ and $\rho$) and nucleons. One of the interesting ideas along these
lines was the effect of the nuclear binding \cite{7,8,9}, which, with ups
and downs, has come to be accepted as, largely or at least partly, responsible
for the depletion of $ R (x) = 2 F_{2 A} (x) / A F_{2 D} (x) $ in the region of
the minimum. 

Criticism on this latter works was raised in \cite{10}, where it was shown 
that  the introduction of relativistic corrections in the usual nuclear 
nonrelativistic treatment of the binding effects resulted in a flux factor
which reduced the conventional binding effects \cite{10,11}. This flux factor
led to a different normalization of the spectral function which preserved
the baryonic number calculated relativistically \cite{10}. The argument of the
normalization of the baryonic number is an important one and the idea has
met with followers \cite{11,12,13}. However, it is a prescription on how 
to convert the nonrelativistic nuclear wave function into a relativistic
spectral function \cite{14} and the prescription is not shared by others
\cite{5}. Furthermore, these are not the only relativistic corrections as
shown in \cite{15,16}.

This issue justifies a work like the present one, where we construct from
the beginning a relativistic nucleon spectral function and define everything
within a field theoretical formalism which uses the nucleon propagators 
written in terms of this spectral function. The relativistic formalism is 
taken from the beginning and the baryonic number is naturally well normalized.
Since all the nuclear information needed is contained in the nucleon spectral
function, one does not need to use ordinary nuclear wave functions, which are
anyway static (no spreading in the energy distribution, which is just 
concentrated in the single particle energies of the shell model) and hence
one does not need any step to introduce relativistic effects into the
nonrelativistic wave functions, as done in \cite{10}. Furthermore, the use
of non static spectral functions is important. This was already seen in refs.
\cite{12,14} which showed that the use of more realistic spectral functions
accounting for nuclear correlations resulted in an enhancement of the binding
effects. The reason is that, for a given average binding energy, the approach 
with a realistic spectral function leads to a larger kinetic and potential
energy in absolute value than the shell model approach.

The other issue that we have revisited here is the pionic contribution. Large
effects from the pionic cloud associated with the pion excess number in the
nucleus were found in \cite{2,3,6}. We have taken up the idea and recalculated
these effects within the many body field theoretical approach using input
which has been checked in a variety of nuclear reactions testing real and
virtual pions: pionic reactions \cite{17}, muon capture \cite{18}, inclusive
neutrino scattering \cite{19} and photonuclear reactions \cite{20}.
In addition we have also included corrections from the modification of 
$\rho$-meson cloud in the nucleus, in complete analogy with the pionic 
contribution.

Altogether we find that the use of the spectral function for the nucleons
together with the mesonic effects can approximately account for the measured 
EMC effect.

The calculations are done using the spectral function for nucleons in nuclear
matter, followed by the local density approximation. This was shown to be
an excellent tool to deal with photonuclear reactions in the absence of
shadowing effects \cite{20}. Hence, a natural limit of our results is the
region of shadowing, $x \leq 0.15$, where indeed there are discrepancies
with the data, and other ingredients should be considered that we do not
want to tackle \cite{21,22}.

\section{Relativistic nucleon propagator in nuclear matter}
\subsection{Nonrelativistic nucleon propagator and spectral functions}
Let us recall first the nonrelativistic nucleon propagator for a noninteracting
Fermi sea, which is given in momentum space by

\begin{equation}
G (p^0, p) = \frac{1 - n (\vec{p})}{p^0 - \varepsilon (\vec{p}) + i \epsilon}
+ \frac{n (\vec{p})}{p^0 - \varepsilon (\vec{p}) - i \epsilon}
\end{equation}

\noindent
where $n (\vec{p})$ is the Fermi occupation number $n (\vec{p}) = 1$ for
$|\vec{p}| \leq k_F , n (\vec{p}) = 0$ for $|\vec{p}| > k_F$ and 
$\varepsilon (\vec{p})$ is the nonrelativistic nucleon energy. Eq. (1) can
be recast as

\begin{equation}
G (p^0, p) = \frac{1}{p^0 - \varepsilon (\vec{p}) + i \epsilon}
+ 2 \pi i n (\vec{p}) \; \; \delta (p^0 - \varepsilon (\vec{p}))
\end{equation}

\noindent
which separates the propagator into the free propagator and the medium
correction.

For an interacting Fermi sea the nucleon propagator can be written in terms 
of its nonstatic selfenergy  $\Sigma (p^0, p)$

\begin{equation}
G (p^0, p) = \frac{1}{p^0 - \varepsilon (\vec{p}) - \Sigma (p^0, p)}
\end{equation}

\noindent
which can be rewritten in terms of the spectral functions for holes and
particles as \cite{23} 

\begin{equation}
G (p^0, p) = \int^{\mu}_{-\infty} 
\frac{S_h (\omega, p)}{p^0 - \omega - i \epsilon} \; \; d \omega +
\int^{\infty}_{\mu} \frac{S_p (\omega, p)}{p^0 - \omega + i \epsilon} d \omega
\end{equation}

\vspace{0.4cm}
\noindent
with the following relationships

$$
\begin{array}{rr}
S_h (p^0, p) = & \frac{1}{\pi} \; \frac{Im \Sigma (p^0, p)}
{[p^0 - \varepsilon (\vec{p}) - Re \Sigma (p^0, p)]^2 +
[Im \Sigma (p^0, p)]^2}\\[2ex]
& \hbox{for} \; p^0 \leq \mu 
\end{array}
$$

\begin{equation}
\begin{array}{rr}
S_p (p^0, p) = & - \frac{1}{\pi} \; \frac{Im \Sigma (p^0, p)}
{[p^0 - \varepsilon (\vec{p}) - Re \Sigma (p^0, p)]^2 +
[Im \Sigma (p^0, p)]^2}\\[2ex]
& \hbox{for} \; p^0 > \mu 
\end{array}
\end{equation}

\vspace{0.4cm}
\noindent
let us also recall that the momentum distribution of the nucleon in this 
interacting Fermi sea is given by

$$
n_I (\vec{p}) = \int^{\mu}_{- \infty} \; S_h (\omega, p) \; d \omega
$$

\begin{equation}
1 - n_I (\vec{p}) = \int^{\infty}_{\mu} \; S_p (\omega, p) \; d \omega
\end{equation}

\noindent
with the automatic sum rule

\begin{equation}
\int^{\mu}_{-\infty} S_h (\omega, p) \; d \omega +
\int^{\infty}_{\mu} S_p (\omega, p) \; d \omega = 1
\end{equation}

\noindent
In passing we also note that in physical reactions $n_I (\vec{p})$ does not
factorize out in the physical cross sections because other factors dependent
on $\omega$ and $\vec{p}$ appear simultaneously in the formulae, and restrictions
due to energy and momentum conservation do not allow the infinite ranges in
the $\omega$ integration required in eq. (6). Failure to realize that, and
the naive substitution of $n (\vec{p})$ in eq. (1) by $n_I (\vec{p})$ of eq. 
(6), as sometimes done, leads to erroneous results which can be off by
three orders of magnitude in some cases \cite{24}. This gives us a 
warning that we should express all our magnitudes in terms of the spectral 
functions, not the momentum distributions.

\subsection{Relativistic nucleon propagator and spectral functions.}

The free relativistic nucleon propagator is given by

\begin{equation}
\frac{\not \! p + M }{p^2 - M^2 + i \epsilon} \equiv
\frac{M}{E (\vec{p})}
\left\{
\frac{\sum_r u_r (\vec{p}) \bar{u}_r (\vec{p})}
{p^0 - E (\vec{p}) + i \epsilon} +
\frac{\sum_r v_r (- \vec{p}) \bar{v}_r (- \vec{p})}
{p^0 + E (\vec{p}) - i \epsilon}
\right\}
\end{equation}

\noindent
where we have separated  in the second member the contribution from positive
and negative energy states \cite{25}.
$M, E (\vec{p})$ in eq. (8) are the nucleon mass and the relativistic nucleon
energy  $(\vec{p}^2 + M^2)^{1/2}$, and $u_r (\vec{p}), v_r (\vec{p})$ are the
ordinary spinors which we take normalized as $\bar{u}_r \, (\vec{p}) \, 
u_r \, (\vec{p}) = 1$.
We recall that $u_r (\vec{p})$ are functions of three momentum and they will
be the only spinors which will appear in our framework.

The relativistic nucleon propagator for a noninterating Fermi sea is easily
derived and, by analogy to eq. (2), can be written as

\begin{equation}
G (p^0, p) = \frac{\not \! p + M}{p^2 - M^2 + i \epsilon}
+ 2 \pi \, i \, n (\vec{p}) (\not \! p + M) \,
\theta \, (p^0) \, \delta \, (p^2 - M^2)
\end{equation}

\noindent
which by means of the identity of eq. (8) can be recast as

$$
G (p^0, p) = \frac{M}{E (\vec{p})} 
\left\{
\sum_r u_r (\vec{p}) \bar{u}_r (\vec{p}) 
\left [ \frac{1 - n (\vec{p})} {p^0 - E (\vec{p}) + i \epsilon} +
\frac{n (\vec{p})}{p^0 - E (\vec{p}) - i \epsilon} \right ] 
\right.
$$

\begin{equation}
\left. + \frac{\sum_r v_r (- \vec{p}) \bar{v}_r (- \vec{p})} 
{p^0 + E (\vec{p}) - i \epsilon}
\right\}
\end{equation}

\noindent
Apart from the negative energy contribution, which will play no role in our
problem, the only difference between eq. (10) and the nonrelativistic 
propagator of eq. (1) is the presence of the factor $M/ E (\vec{p})$ and
the projector over the space of positive energies
$\sum_r u_r (\vec{p}) \bar{u}_r (\vec{p})$,
which are both unity in the nonrelativistic approximation. 

Now we proceed to construct the relativistic propagator in the interacting
Fermi sea. We wish to sum the Dyson series for the diagrams shown in fig. 1,
where although not shown, one would also have other sources of nucleon
selfenergies. We will write them in terms of the operator $\Sigma (p^0, p)$.
It will become clear later on that we only need the imaginary part of the
nucleon propagator for the positive energy states, in which case we neglect
from the beginning the negative energy states (their weight becomes negligible
compared to the singular part of the positive energy propagator). Hence, 
for the purpose of the present problem, the nucleon propagator needed will be

$$
G (p^0, p) = 
\frac{M}{E (\vec{p})} \sum_r u_r (\vec{p}) \bar{u}_r (\vec{p})
\frac{1}{p^0 - E (\vec{p})} +
$$

$$
\frac{M}{E (\vec{p})} 
\sum_r \frac{u_r (\vec{p}) \bar{u}_r (\vec{p})}{p^0 - E (\vec{p})}
\Sigma (p^0, p) \frac{M}{E (\vec{p})} 
\sum_s \frac{u_s (\vec{p}) \bar{u}_s (\vec{p})}{p^0 - E (\vec{p})} + ...
$$

\begin{equation}
= \frac{M}{E (\vec{p})} 
\sum_r \frac{u_r (\vec{p}) \bar{u}_r (\vec{p})}
{p^0 - E (\vec{p}) - \bar{u}_r (\vec{p}) \Sigma (p^0, p) u_r (\vec{p}) 
\frac{M}{E (\vec{p})}}
\end{equation}

\noindent
where we have used the fact that $\Sigma$ should be diagonal in spin for 
spin saturated matter which we only consider.

Comparison of eqs. (11) and (3) shows again the differences between the 
relativistic and nonrelativistic propagators.

The structure of eq. (11) allows one to define a spectral representation
of the nucleon propagator by means of $S_h (\omega, p)$ and $S_p (\omega, p)$ as

$$
G (p^0, p) = \frac{M}{E (\vec{p})} \sum_r u_r (\vec{p}) \bar{u}_r (\vec{p})
\left[
\int^{\mu}_{- \infty} d \, \omega \frac{S_h (\omega, p)}{p^0 - \omega - i \eta}
\right.
$$

\begin{equation}
\left.
+ \int^{\infty}_{\mu} d \, \omega \frac{S_p (\omega, p)}{p^0 - \omega + i \eta}
\right]
\end{equation}

\noindent
with the relationships

$$
S_h (p^0, p) = \frac{1}{\pi} \frac{\frac{M}{E (\vec{p})} I m \Sigma (p^0, p)}
{\left[ p^0 - E (\vec{p}) - \frac{M}{E (\vec{p})} R e \Sigma (p^0, p)
\right] ^2 + 
\left[ \frac{M}{E (\vec{p})} Im \Sigma (p^0, p) \right] ^2}
$$

\hspace{10cm} for $p^0 \leq \mu$

\begin{equation}
S_p (p^0, p) = - \frac{1}{\pi} \frac{\frac{M}{E (\vec{p})} I m \Sigma (p^0, p)}
{\left[ p^0 - E (\vec{p}) - \frac{M}{E (\vec{p})} R e \Sigma (p^0, p)
\right] ^2 + 
\left[ \frac{M}{E (\vec{p})} Im \Sigma (p^0, p) \right] ^2}
\end{equation}

\hspace{10cm} for $p^0 > \mu$

\vspace{0.4cm}
\noindent
where for simplicity $\Sigma$ is now $\bar{u} \Sigma u$ which is independent
of the spin. Eqs. (13) are now the generalizations of eqs. (5) using 
relativistic kinematics. Note that $S_p$ and $S_h$ defined in eq. (12) are 
not the nonrelativistic spectral functions normally used. Hence one should
not expect the same normalization as in \cite{10,11,13}. The normalization 
of $S_h$, which we will need, is easily obtained by imposing baryon number
conservation, as done in \cite{10}. For this purpose we evaluate the electromagnetic
form factor at $q = 0$. For the case of the nucleon (fig. 2a)) we have
(assume all baryons have charge unity for normalization purposes),

\begin{equation}
< N | B^{\mu} | N > \equiv \bar{u} (\vec{p}) \gamma^{\mu} u (\vec{p}) =
B  \frac{p^{\mu}}{M} ; B = 1, \; p^{\mu} \equiv (E (\vec{p}), \vec{p})
\end{equation}

For the case of nucleons in the medium we must evaluate the many body diagram
of fig. 2b).

\begin{equation}
< A | B^\mu | A > = ( - ) \int \frac{d^4 p}{(2 \pi)^4}
V \, i Tr [ G (p^0, p) \gamma^{\mu} ] e^{i p^{0} \eta}
\end{equation}

\noindent
where exp$(i p^0 \eta)$, with $\eta \rightarrow 0^{+}$,  is the convergence
factor for loops appearing at the same time \cite{23} and V the volume of our
normalization box.

By means of eq. (12) we can see that the convergence factor limits the 
contribution to the hole spectral function and we get

$$
< A | B^\mu | A > = V  \int \frac{d^3 p}{(2 \pi)^3} \;
\frac{M}{E (\vec{p})} \; T r \; [ \sum_r u_r (\vec{p}) \bar{u}_r (\vec{p})
\gamma^u ]
$$

$$
. \int^{\mu}_{- \infty} S_h (\omega, p) \; d \omega
$$

$$
= V \int \frac{d^3 p}{(2 \pi)^3} \; \frac{M}{E (\vec{p})} \; 
T r \left[ \frac{( \not \! p + M )_{on \,shell}}{2 M} \, \gamma^u \right]
\int^{\mu}_{- \infty} S_h (\omega, p) \; d \omega
$$

\begin{equation}
= 2 V \int \frac{d^3 p}{(2 \pi)^3} \; \frac{M}{E (\vec{p})} \;
\frac{p^\mu_{on \, shell}}{M} \; \int^{\mu}_{- \infty} \; 
S_h (\omega, p) \; d \omega \,
\equiv B \; \frac{P^{\mu}_{A}}{M_A}
\end{equation}

\noindent
where in the last step we have imposed that this matrix element gives the
right current with B baryons, in analogy to eq. (14), and $P^{\mu}_{A}$ is
the momentum of the nucleus. Note that $p^{\mu}_{on \, shell}$ appears in eq.
(16) because the operator $( \not \! p + M)_{on \, shell}$ comes from 
$u_r (\vec{p}) \bar{u}_r (\vec{p})$ which depends only on $\vec{p}$ 
(it corresponds to free particles with $p^{\mu} \equiv (E (\vec{p}), \vec{p}))$.
Obviously eq. (16) must be evaluated in the rest frame of our Fermi sea where
all magnitudes are defined. Only $\mu = 0$ is then relevant and we
obtain the desired normalization

\begin{equation}
2 V \, \int \frac{d^3 p}{(2 \pi)^3} \; \int^{\mu}_{- \infty} \; 
S_h (\omega, p) \; d \omega = B 
\end{equation}

\noindent
(Note that the factors $\frac{M}{E (\vec{p})}, \frac{p^0_{on \, shell}}
{M}$ have cancelled in eq. (16)). The factor 2 is a spin factor. By a simple
inspection of eq. (17) we can see that with the definition of the spectral
functions in eq. (12) one can use eq. (6) to determine momentum distributions
in both the nonrelativistic and relativistic cases.

In our formalism we do not have a box of constant density, but elements of
volume $d^3 r$ with local density $\rho_p (\vec{r}), \rho_n (\vec{r})$,
the nuclear proton and neutron densities at the point $\vec{r}$.
Hence our spectral functions for protons and neutrons are functions of the
local Fermi momentum 

\begin{equation}
k_{F , p} (\vec{r}) = [ 3 \pi^2 \rho_p (\vec{r}) ]^{1/3} \; ; \;
k_{F , n} (\vec{r}) = [ 3 \pi^2 \rho_n (\vec{r}) ]^{1/3}
\end{equation}

\noindent
and then the equivalent normalization to eq. (17) is

\begin{equation}
2 \int \frac{d^3 p}{(2 \pi)^3} \; \int^{\mu}_{- \infty} \;
S_h (\omega, p, k_{F p,n} (\vec{r})) \; d \omega = \rho_{p, n} (\vec{r})
\end{equation}

\noindent
In practice we shall work with symmetric nuclear matter of density 
$\rho (\vec{r})$. Hence we have a unique Fermi momentum defined as
$k_F (\vec{r}) = [3 \pi^2 \rho (r) / 2]^{1/3}$ and then one has

\begin{equation}
4 \int \frac{d^3 p}{(2 \pi)^3} \; \int^{\mu}_{- \infty} \;
S_h (\omega, p, k_F (\vec{r})) \; d \omega = \rho (\vec{r})
\end{equation}

\noindent
or equivalently

\begin{equation}
\int d^3 r \; \; 4 \int \frac{d^3 p}{(2 \pi)^3} 
\int^{\mu}_{- \infty} \; S_h (\omega, p, k_F (\vec{r})) \; d \omega = A
\end{equation}

\noindent
with $A$ the mass number of each nucleus. The density $\rho (\vec{r})$ of each nucleus
is taken from experiment in our case, and expressed in terms of a two Fermi
parameter distribution for medium and heavy nuclei \cite{26} and mod harmonic 
oscillator for light nuclei \cite{26,27}. Eq. (21) is fulfilled at the level of 
2 - 3 $\%$ in our case, in spite of the non trivial structure of the spectral 
function and the integrals involved. The small numerical deviation
from the right normalization is taken care by dividing by the integral of eq.
(21) instead of by $A$ in the evaluation of $R (x)$, since a similar integration
weighed by the structure functions appears in the numerator, as we shall see.
Although one could separate the contribution
of protons and neutrons in the calculation, we have only applied the results
to nuclei with $N = Z$, or very close, like $^{56} Fe$ and hence, we work
with the symmetric nuclear matter version.

\section{Deep inelastic electron scattering from nuclei}

Let us recall the basic ideas in deep inelastic scattering. Consider the
$(e, e')$ process on a nucleon of fig. 3a). The invariant $T$ matrix for the 
process is 

\begin{equation}
- i T = i e \bar{u}_e (\vec{k'}) \gamma^{\mu} u_e (\vec{k}) \,
\frac{- i g_{\mu \nu}}{q^2} (-i e) < X | J^{\nu} | N >
\end{equation}

\noindent
where $< X | J^{\nu} | N >$ is the invariant matrix element of the 
hadronic current. The cross section for the process $e N \rightarrow e' X$
is given in Mandl and Shaw normalization \cite{28} by 

$$
\sigma = \frac{1}{v_{rel}} \, \frac{2 m}{2 E_e (\vec{k})}\,
\frac{2 M}{2 E (\vec{p})} \, \int \frac{d^3 k'}{(2 \pi)^3} \,
\frac{2 m}{2 E_e (\vec{k}')}
$$

$$
\Pi^N_{i = 1} \int \frac{d^3 p'_i}{(2 \pi)^3} \, \Pi_{l \epsilon f} \,
\left( \frac{2 M'_l}{2 E'_l} \right) \,
\Pi_{j \epsilon b}
\left( \frac{1}{2 \omega'_j} \right)
\bar{\Sigma} \Sigma |T|^2 (2 \pi)^4
$$

\begin{equation}
\delta^4 (p + k - k' - \Sigma^N_{i = 1} p'_i)
\end{equation}

\noindent
where $m$ is the electron mass, $f$ stands for fermions and $b$ for bosons in
the final state X. The factor $\frac{2 M}{2 E (\vec{p})}$ becomes 
$\frac{1}{2 \omega (\vec{p})}$
if we study the cross section on a pion.
The index $i$ is split
in $l, j$ for fermions and bosons respectively.

In the nucleon
rest frame one can then write the differential cross section,
with $\Omega', E'$ referring to the outgoing electron, as

\begin{equation}
\frac{d^2 \sigma}{d \Omega' d E'} = \frac{\alpha^2}{q^4} \; \frac{k'}{k} \;
L'_{\mu \nu} \; W'^{\mu \nu}
\end{equation}

\noindent
with $\alpha = e^2 / 4 \pi$ and $L'_{\mu \nu}$ the leptonic tensor

\begin{equation}
L'_{\mu \nu} = 2 k_{\mu} k'_{\nu} + 2 k'_{\mu} k_{\nu} + q^2 g_{\mu \nu}
\end{equation}

\noindent
and $W'^{\mu \nu}$ the hadronic tensor defined as

\begin{equation}
W'^{\mu \nu} = \frac{1}{2 \pi} \; W^{\mu \nu}
\end{equation}

\noindent
with 

$$
W^{\mu \nu} = \bar{\Sigma}_{s_p} \; \Sigma_X \; \Sigma_{s_i} \Pi^N_{i = 1}
\; \int \frac{d^3 p'_i}{(2 \pi)^3} \; \Pi_{l \epsilon f} \;
\left(
\frac{2 M'_l}{2 E'_l}
\right) \;
\Pi_{j \epsilon b} \;
\left(
\frac{1}{2 \omega'_j}
\right)
$$

\begin{equation}
< X | J^{\mu} | H >^* <X | J^{\nu} | H > (2 \pi)^4 \delta^4 (p + q - 
\Sigma^N_{i = 1} p'_i)
\end{equation}

\noindent
where $q$ is the momentum of the virtual photon, $s_p$ the spin of the
nucleon and $s_i$ the spin of the fermions in $X$. 

Lorentz covariance and gauge invariance allow one to write $W'^{\mu \nu}$ as
\cite{29}

\begin{equation}
W'^{\mu \nu} = 
\left( \frac{q^{\mu} q^{\nu}}{q^2} - g^{\mu \nu} \right) \;
W_1 + \left( p^{\mu} - \frac{p . q}{q^2} \; q^{\mu} \right)
\left( p^{\nu} - \frac{p . q}{q^2} \; q^{\nu} \right)
\frac{W_2}{M^2}
\end{equation}

\noindent
where $W_1, W_2$ are the two structure functions of the nucleon and are
functions of $q^2, p.q $.

Now we evaluate the cross section for $(e, e')$ on the nucleus. In order
not to miss flux factors and be able to write everything in terms of 
propagators, we evaluate the electron selfenergy corresponding to the
diagram in fig. 3 b). We obtain

$$
- i \Sigma (k) = \int \frac{d^4 q}{(2 \pi)^4} \; 
\bar{u_e} (\vec{k}) \; i e \gamma^{\mu} \; i \frac{\not \! k' + m}{k'^{2} -
m^2 + i \epsilon} \; i e \gamma^{\nu} u_e (\vec{k}) \; 
$$

\begin{equation}
\frac{-i g_{\mu \rho}}{q^2} \; (- i) \; \Pi^{\rho \sigma} (q) \; 
\frac{-i g_{\sigma \nu}}{q^2}
\end{equation}

\noindent
which for unpolarized electrons can be written as

\begin{equation}
\Sigma (k) = i e^2 \; \int \frac{d^4 q}{(2 \pi)^4} \;
\frac{1}{q^4} \;
\frac{1}{2m} \;
L'_{\mu \nu} \; \frac{1}{k'^2 - m^2 + i \epsilon} \; \Pi^{\mu \nu} (q)
\end{equation}

\noindent
with $\Pi^{\mu \nu} (q)$ the photon selfenergy.

In order to evaluate the cross section we need only $I m \Sigma (k)$ 
which can be easily evaluated by means of Cutkosky rules \cite{25}.

\begin{equation}
\begin{array}{lll}
\Sigma (k) & \rightarrow & 2 i \; I m \Sigma (k) \\
D (k') & \rightarrow & 2 i \theta (k'^0) \; I m D (k') \; 
\hbox{(boson propagator)} \\
\Pi^{\mu \nu} (q) & \rightarrow & 2 i \theta (q^0) \; I m \Pi^{\mu \nu} (q) \\
G (p) & \rightarrow & 2 i \theta (p^0) \; I m G (p) \; 
\hbox{(fermion propagator)} 
\end{array}
\end{equation}

\noindent
with the result

\begin{equation}
Im \Sigma (k) = e^2 \int \frac{d^3 q}{(2 \pi)^3} \;
\frac{1}{2 E_e (\vec{k} - \vec{q})} \;
\theta (q^0) \; Im \Pi^{\mu \nu} (q) \;
\frac{1}{q^4} \; \frac{1}{2m} \; L'_{\mu \nu}
\end{equation}

\noindent
with $q^0 = k^0 - E_e (\vec{k} - \vec{q})$.

\vspace{0.3cm}
The cross section is readily evaluated from there. Inspection of eq. (11) 
for the relativistic fermion propagator tells us that the electron width is
given by

\begin{equation}
\Gamma (k) = - \frac{2 m}{E_e (\vec{k})} \; I m \Sigma (k)
\end{equation}

\noindent
by means of which the contribution to the cross section from an element of
volume $d^3 r$ in the rest frame of the nucleus is

$$ d \sigma = \Gamma d t d S = \Gamma \frac{dt}{dl} \;
dl dS = \frac{\Gamma}{v} \; d^3 r = $$

\begin{equation}
= \Gamma \; \frac{E_e (\vec{k})}{k} \; d^3 r = - \frac{2 m}{k} \;
Im \; \Sigma \; d^3 r 
\end{equation}

\noindent
Hence we immediately write the $(e, e')$ cross section in the nucleus as

\begin{equation}
\frac{d^2 \sigma}{d \Omega' d E'} = - \frac{\alpha}{q^4} \;
\frac{k'}{k} \; \frac{1}{(2 \pi)^2} \; L'_{\mu \nu} \; 
\int d^3 r \; I m \; \Pi^{\mu \nu} \; (q) 
\end{equation}

\noindent
with $q^0 = k^0 - E_e (\vec{k} - \vec{q}) = k^0 - k'^0$, which is always
positive in this experiment, so we drop the $\theta (q^0)$ function. 
Comparison of eq. (35) with eq. (24) used for nuclear targets tells as that

\begin{equation}
W'^{\mu \nu}_A (q) = - \frac{1}{e^2} \; \frac{1}{\pi} \; \int d^3 r \; 
I m \; \Pi^{\mu \nu} (q) 
\end{equation}

Next we evaluate $\Pi^{\mu \nu} (q)$ corresponding to the right hand side
of the diagram of fig. 3b) using again the Feynman rules in terms of propagators.
We have

$$
- i \Pi^{\mu \nu} (q) = ( - ) \; \int \frac{d^4 p}{(2 \pi)^4} \; i G (p) \;
\Sigma_X \; \Sigma_{s_p, s_i} \Pi^N_{i = 1}
\int \frac{d^4 p'_i}{(2 \pi)^4}
$$

\begin{equation}
\Pi_l i G_l (p'_l) \Pi_j \; i D_j (p'_j) 
( - i )^2 e^2 < X | J^{\mu} | H > < X | J^{\nu} | H >^*
(2 \pi)^4  \; \delta^4 (q + p - \Sigma^N_{i = 1} p'_i)
\end{equation}

\noindent
which by means of Cutkosky rules (31) and the use of free propagators, eq. (8),
for the final states and the medium propagator, eq. (12), for $G (p)$,  
plus eq. (36), leads immediately to

$$
W'^{\mu \nu}_A = \Sigma_{n, p} \int d^3 r \; \int \frac{d^3 p}{(2 \pi)^3} \;
\frac{M}{E (\vec{p})} \; \int ^{\mu}_{- \infty} \; S_h (p^0, p) \, d p^0
$$

$$
\frac{1}{2 \pi} \, \Sigma_{X} \, \Sigma_{s_p} \, \Sigma_{s_i} \, \Pi^N_{i = 1} \,
\int \, \frac{d^3 p'_i}{(2 \pi)^3} \, \Pi_{l \epsilon f} \, 
\left(
\frac{2 M'_l}{2 E'_l}
\right) \, \Pi_{j \epsilon b} \, 
\left( \frac{1}{2 \omega'_j} \right)
$$

\begin{equation}
< X | J^{\mu} | H > < X | J^{\nu} | H >^*
(2 \pi)^4 \delta^4 (q + p - \Sigma^N_{i = 1} p'_i)
\end{equation}

\noindent
which by means of eq. (27) can be rewritten for an isospin symmetric nucleus
as

$$
W'^{\mu \nu}_A = 4 \int \, d^3 r \, \int \frac{d^3 p}{(2 \pi)^3} \, 
\frac{M}{E (\vec{p})} \, \int^{\mu}_{- \infty} d p^0 S_h (p^0, p)
$$

$$ W'^{\mu \nu}_N (p, q) $$

\noindent
with

\begin{equation}
p \equiv (p^0, \vec{p}); \; W'^{\mu \nu}_N = \frac{1}{2} (W'^{\mu \nu}_p +
W'^{\mu \nu}_n) 
\end{equation}

\noindent
Note that $W'^{\mu \nu}_N (p, q)$ appears with the off shell arguments of
$p$, the bound nucleon. 

In the steps from eq.(37) to (38) the spinors $u (\vec{p})$ are included in the
matrix elements of the currents and we have considered that there is necessarily
a fermion loop (hence the first minus sign in eq. (37)) with a free particle
in the final state and the nucleon in the medium in the initial state. The
corresponding energy integration in the loop (if we had used a Wick rotation
explicitly instead of Cutkosky
rules) necessarily picks up the hole part of the propagator of eq. (12), and
at the same time relaxes the condition $\theta (p^0)$ of Cutkosky rules
which does not appear for the hole part. 

In eq. (39) there is an apparent lack of normalization, since assuming
$W'^{\mu \nu}$ constant (which actually cannot be in practice) we would
expect $W'^{\mu \nu}_A = A W'^{\mu \nu}$. However given the normalization
of the spectral function in eq. (21), this is not the case. In eq. (39) we 
get the extra factor $\frac{M}{E (\vec{p})}$ which does not appear in eq.
(21). It is interesting to see the meaning of this factor in eq. (39). If
we look at the formula of the $e N$ cross section in eq. (23), and by means of eqs.
(25) and (27) we find

\begin{equation}
\sigma = \frac{\alpha^2 M}{v_{rel} E_e (\vec{k}) E (\vec{p})} \,
\int \frac{d^3 k'}{E_e (\vec{k}\,^{'})} \,
\frac{1}{q^4} \,
L'_{\mu \nu} W'^{\mu \nu}_N
\end{equation}

\noindent
where $L'_{\mu \nu} W'^{\mu \nu}$ is a Lorentz invariant and the content of the
integral in eq. (40) also. For collinear frames of reference we also have

\begin{equation}
v_{rel} E_e (\vec{k}) E (\vec{p}) = M k
\end{equation}

\noindent
where $k$ is the electron momentum in the frame where the nucleon is at rest
and hence

\begin{equation}
\sigma = \frac{\alpha^2}{k} \, 
\int \frac{d^3 k'}{E_e (\vec{k}\,^{'})} \,
\frac{1}{q^4} \, L'_{\mu \nu} \, W'^{\mu \nu}_N
\end{equation}

Now if we have a system of moving nucleons the cross section for scattering of
the electron with the nucleus can not be obtained as a sum of individual cross
sections, because the relative $e N$ flux is different for each nucleon. 
Instead, one has to sum the probabilities of collision per unit time for each
nucleon and divide by a unique flux, the one relative to the CM of the nucleus.
By taking for $v_{rel}$ in eq. (40) the velocity of the electron with respect to
the CM of the nucleus and suming over all the nucleons in eq. (40) we will
be calculating the electron nucleus cross section. Hence we obtain in the rest
frame of the nucleus

$$
\sigma_A = \frac{\alpha^2}{k} \,
\int \frac{d^3 k'}{E_e (\vec{k}\,^{'})} \, \frac{1}{q^4} \, L'_{\mu \nu} \,
\sum_{\vec{p}} \, \frac{M}{E (\vec{p})} \, W'^{\mu \nu}_N (p, q) 
$$

\begin{equation}
= \frac{\alpha^2}{k} 
\int \frac{d^3 k'}{E_e (\vec{k}\,^{'})} \, \frac{1}{q^4} \, L'_{\mu \nu} \,
4 \int d^3 r \,
\int \frac{d^3 p}{(2 \pi)^3} \, \frac{M}{E (\vec{p})} \,
\int^{\mu}_{- \infty} d p^0 S_h (p^0, p) \,
W'^{\mu \nu}_N (p, q)
\end{equation}

\noindent
where $k$ is the electron momentum in the nucleus rest frame. Since our 
nuclear cross section is given in terms of $W'^{\mu \nu}_A$ by (see eqs.
(35), (36))

\begin{equation}
\sigma_A = \frac{\alpha^2}{k} \, \int \frac{d^3 k'}{E_e (\vec{k}\,^{'})} \,
L'_{\mu \nu} \, W'^{\mu \nu}_A
\end{equation}

\noindent
then eq. (39) follows immediately.

The previous discussion has shown that the factor $\frac{M}{E (\vec{p})}$
is a factor appearing in the probability of reaction per unit time for each
nucleon, and remains in the integral when we divide by a unique flux in order
to obtain the nuclear cross section. It is thus a Lorentz contraction factor.

In the limit of small densities, when $M / E (\vec {p}) = 1$, eq. (39) with
the consideration of eq. (21) would give $W'^{\mu \nu}_A = A W'^{\mu \nu}_N$
as it should be. Eq. (39) accounts for Fermi motion and binding and includes 
the relativistic Lorentz contraction
factor $M / E (\vec{p})$ and the change of the arguments in 
$W'^{\mu \nu} (p, q)$. Note in passing that the relativistic factor
$m / E_e (\vec{k})$ of eq. (33), which we extracted from the relativistic 
propagator of eq. (11), has been essential to provide the right normalization.

\section{Contribution from the pion cloud}

Let us first see the free pion structure function. The same formula eq. (24) 
is  used for pions and this defines $W'^{\mu \nu}_{\pi}$. Given the normalization
of the fields which we follow \cite{28}, the cross section of eq. (23) contains
the factor $1 / 2 \omega (\vec{p})$ instead of $2 M / 2 E (\vec{p})$. Hence,
this means that the definition $W'^{\mu \nu}_{\pi}$ is given by eq. (27) 
dividing the right hand side of the equation by $2 m_{\pi}$ (and obviously 
the average over the spin of the
nucleon does not appear now for the pion case).

In order to derive the contribution from the virtual pions in the medium we
evaluate again the electron selfenergy related to the diagram of fig. 4.
We can save all the steps given before simply by noting the differences in
the two cases:

i) The bound nucleon propagator is substituted by a pion propagator. From
the use of Cutkosky rules we must change

$$- 2 \pi \, \frac{M}{E (\vec{p})} \,
\int^{\mu}_{- \infty} \, d \omega S_h (\omega, p) \, \delta \, (p^0 - \omega)$$

by

\begin{equation}
2 \theta (p^0) \, I m \, D (p)
\end{equation}

\noindent
with $D (p)$ the pion propagator (in the medium).

ii) One must take into account that $W'^{\mu \nu}_{\pi}$ is given by eq. (27)
divided by $2 m_{\pi}$.

iii) The sum over spins of the bound nucleon in eq. (37) does not appear now
for the case of the pion.

iv) There are three charged states of pions.

With only these four rules we can already write

\begin{equation}
W'^{\mu \nu}_{A, \pi} = 3 \int d^3 r \; \int \frac{d^4 p}{(2 \pi)^4} \;
\theta (p^0) (- 2) \; Im D (p) \; 2 m_\pi W'^{\mu \nu}_{\pi} (p, q)
\end{equation}

Now there are two obvious
subtractions to eq. (46). First one should subtract the contribution from
a free pion, which has nothing to do with medium effects. However, this is
zero because one electron can not decay into another electron, one pion and
X. Then assuming the pion is dressed in the medium by exciting $p h $ and
$\Delta h$ as we shall do, one is left with the contributions shown in fig. 5.

Now there is no problem to get a contribution since $Im D (p)$ gets strength
from $ph$ excitation. The physical channel would correspond to
$e \rightarrow e' + X + ph$, or equivalently $e N \rightarrow e' N' X$, 
which is now allowed. The physical channels are easily visualized by cutting
the intermediate states in the diagramas with a horizontal line and placing
on shell the particles cut by the line. This is actually the essence of 
Cutkosky rules to obtain the imaginary part of the selfenergy of a diagram.

The former discussion also tells us that part of what we are calculating is 
already contained in the nucleon structure function. This is because we are
also calculating the contribution from the pions contained in a free nucleon.
This has to be subtracted. This is easily done by substituting in eq. (46)

\begin{equation}
Im D (p) \; \rightarrow \; \delta I m D (p) \equiv I m D (p) - \rho \;
\frac{\partial Im D (p)}{\partial \rho} \left|_{\rho = 0} \right.
\end{equation}

\noindent
since we substract $A$ times the contribution from the pion cloud to the
structure function of the free nucleon. In technical words, we can say that
we are only considering terms with at least two $ph$ or $1 ph 1 \Delta h$ (the most important
terms) etc. in fig. 5 (up to Pauli blocking corrections in the $ph$ Lindhard
function, which are automatically included by the procedure of eq. (47)).

Hence the genuine pionic contribution is given by

\begin{equation}
W'^{\mu \nu}_{A, \pi} = 3 \int d^3 r \; \int \frac{d^4 p}{(2 \pi)^4} \;
\theta (p^0) \; (- 2) \delta Im D (p) \; 2 m_{\pi} \; W'^{\mu \nu}_{\pi} \;
(p, q) 
\end{equation}

\noindent
Eqn. (39) and (48) are the basic equations which provide the nucleonic and
pionic contributions.

In passing we can mention that the distribution of the excess number of pions,
per unit volume in the nucleus, $\delta N_{\pi} (p)$, 
which contains the averages of $<a^+_p a_p>$, $<a^+_p a^+_{- p}>$ and
$<a_p  a_{-p}>$, is given by \cite{30}

\begin{equation}
\frac{\delta N_{\pi} (\vec{p})}{2 \omega (\vec{p})} = - 3 \int^{\infty}_0 \;
\frac{d p^0}{2 \pi} \; \delta I m D (p)
\end{equation}

\noindent
such that in the case of a structure function $W'^{\mu \nu}_{\pi } (p, q)$
independent of $p^0$, eq. (48) could be considered as a weighed integral of
the pion structure function over the pion excess distribution in the nucleus.
However, the strong dependence of $W'^{\mu \nu}_{\pi} (p, q)$ on $p^0$
(imposed by energy and momentum conservation) does not allow that factorization,
and hence a relationship of the pion excess number with the pionic contribution 
to the structure function cannot be established. The apparent extra factor 2
which we obtain in this counting (apart from the Lorentz contraction factor,
$m_{\pi} /\omega (\vec{p}))$
is obtained because one is automatically accounting for the 
imaginary part of the Compton $\gamma \pi$ amplitude which is crossing 
symmetric and contains the two diagrams of fig. 6, while the pion structure
function for on shell pions contains only the imaginary part of the diagram
6a). (see refs. \cite{30} and \cite{31} for an elaborate discussion of these
issues in the problem of the pion cloud contribution to $K^+$ nucleus scattering).
It is worth noting that there is no overlap between the Feynman diagrams 
accounted for in the pionic contribution, fig 5, and those of the nucleonic
contribution with Fermi motion and binding, which come from selfenergy 
insertions in the nucleon line of fig. 3b). Hence, these contributions to the
nuclear structure function are independent.

\section{The Bjorken limit}

We have evaluated the contribution of nucleons and pions to the hadron
structure function of the nucleus. We now proceed to write these expressions
in terms of the Bjorken structure functions \cite{32}.

For nucleons (and similarly for pions or the nucleus) one introduces 
the Bjorken variables 

\begin{equation}
x = \frac{- q^2}{2 p q} \; ; \; \nu = \frac{p . q}{M} \; ; \; Q^2 = - q^2
\end{equation}

\noindent
and for large values of $q^0$ and $Q^2$ simultaneously and fixed $x$ one has
the Bjorken scaling

$$
\nu  W_2 (x, Q^2)  \equiv F_2 (x) 
$$

\begin{equation}
M  W_1 (x, Q^2)  \equiv F_1 (x)
\end{equation}

\noindent
and the Callan-Gross relation

\begin{equation}
2 x F_1 (x) = F_2 (x)
\end{equation}

\noindent
up to some, $Q C D$ corrections in $l n Q^2$. Since the same $Q^2$ will be chosen
for the nucleus and the nucleon and we perfom ratios of structure functions
we shall not worry about this dependence here.

In view of the relations (51), (52) the most practical way to proceed is to
work with transverse components of $W'^{\mu \nu}$. For this purpose assume
$\vec{q}$ along the $z$ direction, as usually done in the study of the $e, e'$
reaction, and evaluate $W'^{xx}$. We find from eq. (28)

$$ W'^{xx} \; = \; W_1 \; + \; \frac{(p_x)^2}{M^2} \; W_2 \; \equiv \; 
\frac{F_1 (x)}{M} \; + \;
\frac{(p_x)^2}{M^2} \; \frac{F_2 (x)}{\nu} $$
 
\begin{equation}
\hspace{6.3cm} = \frac{F_1 (x)}{M} \quad \hbox{in the Bjorken limit}
\end{equation}

\noindent
this component has the virtue that the coefficient of $W_1$ is independent
of $p$ and hence is the same for on shell or off shell nucleons, or pions, or
the nucleus. Hence we can write

$$ \frac{F_{1 A, N} (x_A)}{M_A} = 4 \int d^3 r \; \int \frac{d^3 p}{(2 \pi)^3} 
\; \frac{M}{E (\vec{p})} \; \int^{\mu}_{- \infty} \; d p^0 S_h (p^0, p)
\; \frac{F_{1 N} (x_N)}{M} $$

$$ \frac{F_{1 A, \pi} (x_A)}{M_A} = 3 \int d^3 r \; \int \frac{d^4 p}{(2 \pi)^4} 
\; \theta (p^0) (- 2) \; \delta Im D (p) \; 2 m_{\pi} 
\; \frac{F_{1 \pi} (x_\pi)}{m_{\pi}} $$

$$ x_A = - q^2 / 2 M_A q^0 \equiv \frac{x}{A} \quad , \quad \hbox{with} \; 
x = - q^2 / 2 M q^0 $$

\begin{equation}
x_N = - q^2 / 2 p q \quad \quad ; \quad \quad  x_{\pi} =  - q^2 / - 2 p q
\end{equation}

\noindent
where the extra minus sign in $x_{\pi}$ is because of the direction of $p$ in 
fig. 4.

Here $F_{1 N} (x) = (F_{1 p} (x) + F_{1 n} (x)) / 2 $ as implicit
in eq. (39).

We still have to exert some caution since $0 < x_N < 1 $. On the other hand
$x_{\pi} < 1 $ but $x_{\pi} > x $ because in our scheme the emerging particle
from the coupling of the pion to a nucleon of the nucleus is on shell when 
we excite $p h, \Delta h$ with the pion and take the imaginary part of $D (p)$
\cite{3}.

Since usually one compares ratios of the $F_2$ structure functions, this is
easily accomplished by making use of the Callan-Gross relation (52) and we
find 

\begin{equation}
F_{2 A, N} (x_A) = 4 \int  d^3 r  \int  \frac{d^3 p}{(2 \pi)^3} \; 
\frac{M}{E (\vec{p})} \; \int ^{\mu}_{- \infty} \; d p^0 S_h (p^0, p)
\; \frac{x}{x_N} \; F_{2 N} (x_N) \; \theta (x_N) \; \theta (1 - x_N)
\end{equation}

\begin{equation} 
F_{2 A, \pi} (x_A) = - 6 \int  d^3 r  \int  \frac{d^4 p}{(2 \pi)^4} \; 
\theta (p^0) \; \delta I m D (p) \; 
\; \frac{x}{x_\pi} \; 2 M \; F_{2 \pi} (x_\pi) \; \theta (x_\pi - x) \; 
\theta (1 - x_\pi) 
\end{equation}

\noindent
where we have again $F_{2 N} = (F_{2 p} + F_{2 n}) / 2$ as implicit in eq. (39).

In the Bjorken limit the evaluation of eq. (55) does not require the knowledge
of the variable $q$ since

\begin{equation}
\frac{x}{x_N} \rightarrow \frac{p^0 - p^3}{M} \quad ; \quad 
\frac{x}{x_{\pi}} = \frac{- p^0 + p^3}{M}
\end{equation}

\noindent
but it is implicit in the structure functions which are taken at a certain 
$Q^2$. 

When reaching this point it is worth considering also the contribution of
the $\rho$ meson cloud. Both the pion and the $\rho$ meson couple to nucleons
and delta with derivative couplings, which give rise to relatively large meson
selfenergies in the range of momenta which contributes to the structure functions.
Furthermore, as found already in ref. \cite{30}, the pion cloud contribution
comes mostly from the combined $ph$ and $\Delta h$ excitation, with the 
$ph$ on shell (in $\delta Im D (p))$, and one finds a negligible contribution
of two $ph$ excitations. Only $\pi$ and $\rho$ can excite the $\Delta h$
components and this makes these two mesons special when looking at the
mesonic contribution to the nuclear structure function. In addition there
is an interplay between $\pi$ and $\rho$ exchange. 
Indeed, the $\rho$ meson, through nuclear correlations, contributes 
both to the longitudinal and transverse parts of the spin-isospin $ph$ and
$\Delta h$ interaction and it is an important element contributing to the
Landau-Migdal $g'$ parameter in a microscopic derivation of this interaction
\cite{33}. The value of this parameter governs to some extend the pionic
(and $\rho$-meson) contribution to the structure function \cite{3}.

The contribution of the $\rho$-meson cloud to the structure function is given,
by analogy to eq. (56) by

$$
F_{2 A, \rho} (x_A) = - 12 \int d^3 r \int \frac{d^4 p}{(2 \pi)^4}
\theta (p^0) \delta Im D_{\rho} (p) \frac{x}{x_{\rho}} \, 2 M
F_{2 \rho} (x_{\rho})
$$

\begin{equation}
\hspace{6cm} \theta (x_{\rho} - x) \theta (1 - x_{\rho})
\end{equation}

\noindent
where $D_{\rho} (p)$ is now the $\rho$-meson propagator and $F_{2 \rho}
(x_{\rho}) $is the $\rho$-meson structure function, which we take equal to the 
one of the pion following refs. \cite{34,35}. In addition $x_{\rho}$ is also 
given, in analogy to eq. (57), by

\begin{equation}
\frac{x}{x_{\rho}} = \frac{- p^0 + p^3}{M}
\end{equation}

Eq. (58) contains an extra factor of two compared to the pionic contribution
of eq. (56). This is because of the two transverse polarizations of the
$\rho$-meson and the fact that the coupling of the $\rho$ to nucleons and
deltas which we consider, following ref. \cite{33} is only of transverse
nature $((\vec{\sigma} \times \vec{p}) \vec{\epsilon}$ for nucleons and
$(\vec{S} \, ^{+} \times \vec{p}) \vec{\epsilon}$ for deltas, with $\vec{\epsilon}$ the
polarization vector of the $\rho$ meson and $\vec{S} \, ^{+}$ the spin 
transition operator from spin $1/2$ to $3/2$).

The expression of $F_{1A,N} (x_A)$ in eq. (54) shows the same lack of 
normalization discussed in connection with eq. (39), since assuming
$F_{1 N} (x_N)$ constant (which is not the case) $F_{1A,N}$ is not $A$
times $F_{1N}$, due to the Lorentz contraction factor $\frac{M}{E(\vec{p})}$.
The same can be said about eq. (55) in general. However, eq. (55), shows a 
particular normalization property. Indeed, if we take an ensemble of nucleons
on shell $(p^0 = E (\vec{p}))$ and $x = 0$ (and hence $x_N = 0$, by virtue
of eq. (57)), then the Lorentz contraction factor $\frac{M}{E (\vec{p}))}$
cancells in average the dynamical factor $\frac{x}{x_N}$ and we get
$F_{2A,N} (0) = A F_{2N} (0)$ for on shell nucleons. This cancellation, 
however, will not show up when the nucleons are off shell since in our 
formalism we still obtain the factor $\frac{M}{E (\vec{p})}$ but the dynamical
factor $\frac{x}{x_N}$ will now be different.

In this discussion we are implicitly assuming that eqs. (54) and (55) stand 
as they are for the case of off shell nucleons simply by taking for $x_N$ the
expression of eq. (57) using the nucleon off shell variables. This is 
certainly the easiest form of the analytical continuation in the off shell
regime, although there have been other prescriptions in the Literature 
\cite{36,37,38}.

We would like to justify our assumption. Indeed, in our frawework we use
nucleon propagators which are based on the free spinors (eq. (12)). Hence
in eq. (38) the matrix elements would be defined in terms of free spinors,
the final particles are free particles and the off shell dependence appears
only in the $\delta ( \, )$ function. Obviously one can not look in detail
at all channels implicit in eq. (38). However, one can use the same philosophy
in the parton model which is used to find out the scaling of the structure
functions. We follow here the steps of ref. \cite{32} and assume that partons
carry a fraction $x_q$ of the nucleon momentum $p$ and its mass $M$, that
the electron parton amplitude is given by the on shell expression and that
the outgoing parton is a free one, but the  $\delta ( \, )$ function appears
with the off shell variables.

One finds then

$$
W_1 = W^{'x x} = \sum_i \, \int d x_q f_i (x_q) e^2_i \frac{1}{2} (- q^2)
\frac{1}{x_q M} \delta (q^2 + 2 p q x_q)
$$

\begin{equation}
\; \; \; \; \quad = \sum_i e^2_i f_i (x_N) \frac{1}{2 M} = \frac{F_1 (x_N)}{M} ;
x_N = \frac{- q^2}{2 p q}
\end{equation}

\noindent
where $F_1$ appears whith argument $x_N$ which is defined in terms of the
off shell variables. Similarly one obtains

\begin{equation}
\frac{p q}{M} W_2 = \sum_i e^2_i f_i (x_N) x_N = F_2 (x_N)
\end{equation}

\noindent
whith $p.q$, and $x_N$ defined in terms of the off shell variables.

We would like to note here that our finding $F_{2A} (0) \neq A F_{2N} (0)$,
or equivalently $R (0) \neq 1$,
where $R (x) = F_{2A} (x) / A F_{2N}(x)$,
is not so unconventional. Indeed, in ref.
\cite{15} the authors also find this property and they literally quote
``Note that a feature of the results is that $R (0) \neq 1$...  It does not
reflect any violation of baryon conservation, which is ensured by the
normalization condition", in our case eq. (21).

The issue of the normalization still stirs much controversy. We have devoted
many thoughts to it throughout this paper looking at it from some points 
of view not discussed
before. We see that for an ensemble of uncorrelated nucleons $F_{2 A, N} (0)=
A F_{2 N} (0)$ but as soon as interactions are accounted for this 
normalization is lost. A similar thing would happen should we evaluate the 
structure function $F_{3 A}$, which appears in neutrino scattering. This
structure function when integrated over $x$ is normalized to 3 for the nucleon
if one assumes models with only three valence quarks, or to $3 A$ for the 
nucleus.
Once again this normalization would be lost if QCD interaction corrections are
accounted for \cite{100}
and equivalently if NN interactions are included in the nuclear case, and
experimentally this is the case \cite{100}.
Also, experimentally $F_{2 A, N} (0)\neq A F_{2 N} (0)$. Clear as the      
question looks to us, we are aware that this point of view is not universally
accepted. Further thoughts and discussions on the issue should be welcome
which would help settle the question in a way acceptable to all.

Eqs. (55, 56, 58) are the final equations which we use in the analysis.

For the nucleon and pion structure functions $F_{2N} (x), F_{2 \pi} (x)$ we 
take the experimental values of refs. \cite{39,40}.

\section{The meson propagators}

The pion propagator in the medium is given by

\begin{equation}
D (p) = [ p^{0 2} - \vec{p}\,^{2} - m^2_{\pi} - \Pi_{\pi} (p^0, p) ]^{- 1}
\end{equation}

\noindent
with $\Pi_{\pi}$ the pion selfenergy. We consider the contribution of the $p h$
and $\Delta h$ excitations to the pion selfenergy in connection with the 
Landau-Migdal correction in the pionic channel, as well as off shell pion
nucleon form factors. 

Since for $\delta Im D$ we need $D - D_0$, it is practical to write

\begin{equation}
D (p) - D_0 (p) = D^2_0 (p) \;
\frac{ \frac{f^2}{m_{\pi^2}} F^2 (p) \vec{p} \, ^{2} \, \Pi^* (p)}
{1 - \frac{f^2}{m^2_{\pi}} V'_L (p) \, \Pi^* (p)}
\end{equation}

\noindent
where $F (p)$ is the $\pi N N$ form factor, which we take of the monopole
type

\begin{equation}
F (p) = \frac{\Lambda^2 - m_{\pi^2}}{\Lambda^2 - p^2}
\end{equation}

\noindent
with $\Lambda = 1300 \, MeV$ \cite{41} and $f^2 /4 \pi = 0.08$. 
$V'_L (p)$ is the longitudinal part of the spin-isospin interaction and
$\Pi^* (p)$ is the irreducible pion selfenergy, which contains all 
selfenergy diagrams which are not connected by $V'_L (p)$.

For the $\rho$-meson we can write

\begin{equation}
D_{\rho} (p) - D_{0 \rho} (p) = D^2_{0 \rho} (p) \;
\frac{\frac{f^2}{m_{\pi^2}} C_{\rho} F^2_{\rho} (p) \vec{p} \, ^{2} \, 
\Pi^* (p)}
{1 - \frac{f^2}{m^2_{\pi}} V'_T (p) \, \Pi^* (p)}
\end{equation}

\noindent
with $V'_T (p)$ the transverse part of the spin-isospin interaction, 
$C_{\rho} = 3.94$ \cite{41} and $F_{\rho} (p)$ the $\rho N N$ form factor 
given by

\begin{equation}
F_{\rho} (p) = \frac{\Lambda_{\rho}^2 - m^2_{\rho}}{\Lambda^2_{\rho} - p^2}
\end{equation}

\noindent
and $\Lambda_{\rho} = 1400 MeV$ \cite{41}.

For $V'_L (p), V'_T (p)$ we take the expressions which are derived from a
model with $\pi$ and $\rho$ exchange in the presence of short range nuclear
correlations \cite{42}, which are given by

$$
V'_L (p) = \vec{p} \, ^{2} D_0 (p) F^2 (p) - \vec{p} \, ^{2} \tilde{D}_0
(p) \tilde{F}^2 (p)
$$

\begin{equation}
- \frac{1}{3} q^2_c \tilde{D}_0 (p) \tilde{F}^2 (p) 
- \frac{2}{3} q^2_c \tilde{D}_{0 \rho} (p) \tilde{F}^2_{\rho} (p) C_{\rho}
\end{equation}

$$
V'_T (p) = \vec{p} \, ^{2} D_{0 \rho} (p) F^2_{\rho} (p) C_{\rho}
- \frac{1}{3} q^2_c \tilde{D}_0 (p) \tilde{F}^2 (p) 
$$

\begin{equation}
- (\vec{p} \, ^{2} + \frac{2}{3} q^2_c)
 \tilde{D}_{0 \rho} (p) \tilde{F}^2_{\rho} (p) C_{\rho}
\end{equation}

Here $q_c \simeq 780 MeV$ is the inverse of a typical correlation distance
and $\tilde{D} (p), \tilde{F} (p), \tilde{D}_{\rho} (p), \tilde{F}_{\rho} (p)$
are the corresponding propagators and form factors substituting
$\vec{p} \, ^{2}$ by $\vec{p} \, ^{2} + q^2_c$.
The irreducible selfenergy $\Pi^* (p)$ is in our case the sum of the Lindhard
functions $U_N (p), U_{\Delta} (p)$ for $ph$ and $\Delta h$ excitation
with the normalization and analytical expressions of the appendix
of ref. \cite{43}. It is interesting to note that for the values of $p^0, p$
which contribute most to the structure function, both $V'_L (p)$ and
$V'_T (p)$ are negative and this leads to positive values of $F_{2A,\pi}$
and $F_{2A,\rho}$.

It was also found in \cite{30} that in order to evaluate the contribution of
the pion cloud by using the pion propagator, as done here, it is important
that it satisfies the sum rule

\begin{equation}
- \int^{\infty}_0 \; \frac{d p^0}{\pi} \; 2 p^0 \; I m D (p^0, p) = 1
\end{equation}

\noindent
which expresses the equal time commutation relation of the pion fields. Our
model satisfies this equation at the level of one per thousand.

\section{The nucleon spectral function}

Section 2.2 has established the framework for the relativistic nucleon propagator
which we need here. The only input needed is the nucleon selfenergy. We take
it from the work of ref. \cite{44}. This is a semiphenomenological, quite
succesful approach, which uses as input the $NN$ cross section and the 
spin-isospin effective interaction. This allows one to evaluate $Im \Sigma$, 
which is remarkably close to $Im \Sigma$ of the elaborate many body calculations
of ref. \cite{45,46}. The real part is evaluated by means of a dispersion
relation, and the Fock term from the pionic contribution is also included. Only
pieces of the Hartree type, which should be independent of the momentum,
are missing for which one needs more information.
Hence, up to an unknown momentum
independent piece in the selfenergy the rest of the nucleon properties in the
medium can be calculated, like effective masses, spectral functions, etc, 
which are also in good agreement with sofisticated many body calculations
\cite{47,48}. Actually, what might appear a drawback is now a welcome feature
because since the proper binding energy is an important ingredient in the 
$EMC$ effect, we also include phenomenologically a function $C (\rho)$ in the nucleon
selfenergy and demand that the binding energy per nucleon be the experimental
one for each nucleus. Then the model is complete, realistic and technically
much simpler to handle than the sofisticated many body calculations 
\cite{45,47,48}.

With this improvement, momentum distributions and average binding energies
are also in good agreement with other infinite nuclear matter \cite{49} and
finite nuclei calculations \cite{50}.

A small inconvenience appears because the selfenergy of ref. \cite{44} is
evaluated non relativistically. However we have checked that a proper
calculation including relativistic factors of the type $M/ E (\vec{p})$ in
the nucleon propagators in the integrals over the loops which appear in the
evaluation of $I m \Sigma$ in \cite{44}, would only introduce corrections
in  $I m \Sigma$ below the level of $10 {\%}$. Second, we have changed $I m \Sigma$ 
by $10 {\%}$ and found that the ratio $R (x)$ changes only at the level of 
$1 {\%}$. Thus we take the values for $\Sigma$ from ref. \cite{44} and use 
them in the relativistic propagators of section 2.2. For the average kinetic 
and total nucleon energy we have

$$
< T > =  \frac{4}{A} \int d^3 r \; \int \frac{d^3 p}{(2 \pi)^3} \;
(E (\vec{p}) - M) \; \int^{\mu}_{- \infty} S_h (p^0, p) d p^0 
$$

\begin{equation}
\hspace{-2cm}
< E > =  \frac{4}{A} \int d^3 r \; \int \frac{d^3 p}{(2 \pi)^3} \;
 \int^{\mu}_{- \infty} S_h (p^0, p) p^0 \, d p^0 
\end{equation}

\noindent
and the binding energy per nucleon is then given by the sum rule \cite{51}

\begin{equation}
| E_A | = - \frac{1}{2} \left( < E - M > + \frac{A - 1}{A - 2} < T > \right)
\end{equation}

\noindent
which is also used in \cite{52,53} in connection with the study of the
$EMC$ effect. We take experimental numbers for each nucleus for $| E_A |$
and adjust the function $C (\rho)$ to fit $| E_A |$. We take $C (\rho)$
linear in the density, $C \rho (r)$. This quantity provides around $30 MeV$ 
repulsion at $\rho = \rho_0$ in most of the nuclei. Detailed values 
for $<T>, <E>$ and $E_A$ can be seen in Table I.

As noted in refs. \cite{12,52,53}, the use of nucleon propagators in terms of
non static spectral functions leads to bigger values of the average kinetic
energy and $| < E - M > |$ than the shell model of the nucleus and as a 
consequence to reduced values of $R (x)$ (for $x < 0.7$).
We can see this here also by taking the uncorrelated Fermi sea and adding 
a function $D \rho (r)$ to the ordinary Thomas-Fermi potential 
$V_{T F} (r) = - k_F (r)^2 / 2 M $ such as to get the same binding energy via
eq. (71). We can make use of the same formalism by simply considering that

$$
S^{U F S}_h (p^0, p) = n (\vec{p}) \; \delta (p^0 - E (\vec{p}) - \Sigma)
$$

\begin{equation}
\Sigma (r) = V_{T F} (r) + D \rho (r)
\end{equation}

Eq. (72) associates one energy to a given momentum, (the essence of the shell
model in infinite nuclear matter) while the spectral function has a peak
around the quasiparticle energy and then spreads out at larger values of
$\vec{p}$ for a given energy. This results in a larger value of the average
kinetic energy.

In Table I we show the results for different nuclei and compare them with 
those of ref. \cite{52}. As one can see, the results that we obtain with
the uncorrelated Fermi sea and the spectral functions are remarkably close
respectively to those of the Hartree Fock and spectral function used in
\cite{52}.

We shall evaluate $R (x)$ using both the spectral function approach and
the uncorrelated Fermi sea. We have integrated over the momentum up to
four times the Fermi momentum for each energy. This gives the normalization
of $A$ at the level of $2 - 3 \%$ which is sufficient for our purposes, but
can lead to higher uncertainties in the kinetic energy, which  weighs the
integral with a higher power of $\vec{p}$. Even a conservative error of $20 \%$
in the kinetic energy has repercussions in the $E M C$ effect only at the 
level of $2 \%$.

\section{Results and discussion}

\subsection{Nucleonic contribution}

In fig. 7 we show the results for $R (x)$ from the nucleonic contribution
calculated as

\begin{equation}
R_N (x) = \frac{F_{2 A, N} (x_A)}{A F_{2 N} (x)}
\end{equation}

We do not divide by $F_{2D} (x)$ as experimentally done. One reason for it is
that the techniques used here with the local density approximation cannot be
used for deuterium and hence we cannot calculate the nuclei and deuterium
with the same model. However the price we pay is small. We rely upon the
results for $F_{2D} (x)$ calculated for deuterium in \cite{12}. We can see
there that for $0 < x < 0.7 \quad F_{2D} (x) / F_{2N} (x)$ ranges between $0.98 - 1$.
Hence, our results should be increased by $ 1 - 2 \%$ and we will not worry
about this amount. More serious is the region for $x > 0.7$, where due to
Fermi motion the former ratio increases rapidly. Hence, in that region we should
expect to overcount the experiment as it is indeed the case. 

In fig. 7 we plot the results obtained for $^{56} Fe$.
This nucleus has $N \neq Z$ but by a little amount. Furthermore,
the experimental
results are corrected by the isoscalarity factor to convert them into an
equivalent isoscalar nucleus \cite{12,54}, hence our calculations done for symmetric
nuclear matter are appropriate. We see a minimum around
$x = 0.6$ as in the experiment and a steep rise at $x > 0.75$ as it correponds
to Fermi motion \cite{55,2,5}. The region around $x = 0.6$ agrees well with
experiment. However at $x \simeq 0.15 - 0.2$ we are below the data by about
$10 - 15 \%$. This region will be filled up latter by the mesonic contribution.

In the same figure we show the results obtained with the uncorrelated Fermi
sea (local step function distribution) of eq. (72). We observe that $R_N (x)$ 
takes values closer to unity than the results with the spectral function
for $x < 0.6$
The reduction of $R_N (x)$ with the use of the spectral function with respect
to a static picture of the nucleus, like Hartree Fock or the equivalent
uncorrelated Fermi sea in our case, was already shown and explained in ref.
\cite{52}. The results here are qualitatively similar to those in \cite{52}
and the explanation lies in the increased binding provided by the spectral
functions. Both the factor $x / x_N$ of eq. (57), as well as the restrictions
of phase space $\theta (x_N) \theta (1 - x_N)$, are responsible for the decrease
of $R_N (x)$ in this region.

On the other hand there is a novelty in these results with respect to those
in \cite{52}. $R_N (x)$ does not go to 1 at $x = 0$ as in \cite{52} and are
systematically lower in all the range of $x$. The reduction at $x = 0$ is easy
to see from the formulae. In eq. (55) $F_{2 N} (x_N = 0)$ will take a constant
value, and with respect to the normalization integral of eq. (21) the novelties 
in eq. (55) are the
extra factors $M / E (\vec{p})$ and $x / x_N$ which both go into reducing
the contribution of the integrand for off shell nucleons.

The overshooting of the results in the region of $\; x \simeq 0.8$ was already
announced as a result of dividing $F_{2 A}$ by $F_{2 N}$ and not $F_{2D} /2$. But
the qualitative features due to Fermi motion are reproduced. A very detailed
discussion of these effects is given in ref. \cite{5}, but qualitatively 
we can see that

$$ x_N \rightarrow \frac{x_N}{x} = \frac{M}{p^0 - p^3} $$

\begin{equation}
\hspace{-2cm}
(x \rightarrow 1)
\end{equation}

\noindent
and one can pick up values of $p^3$ in the integrations such that $x_N < 1$.
Hence $F_{2 A, N}$ will be different from zero while $F_{2 N} (x = 1) = 0$, and
$R_N (x)$ from eq. (73) necessarily goes to $\infty$.

It is interesting to call on the attention to the crossing of the two lines
in fig. 7. The binding effects reduce $R_N(x)$. On the other hand Fermi motion
increases $R_N(x)$ close to 1, as we noted. Fermi motion is more important in
the interacting Fermi sea because now one has larger momentum components. On
the other hand the interacting Fermi sea has also more binding. As a consequence
we see that at $x < 0.6$, where the binding effects dominate over the Fermi
motion, $R_N(x)$ for the interacting Fermi sea is smaller than with the non
interacting Fermi sea, while for $x > 0.7$, where the Fermi motion effect
dominates, the situation is just opposite.

\subsection{Mesonic contributions}

In figs. 8, 9, 10, 11 we show the mesonic contribution to $R (x)$, 
together with the nucleonic one discussed above, for different nuclei,
$^{6}Li$, $^{12}C$, $^{40}Ca$ and $^{56}Fe$. The general features are the same
in all nuclei, but, of course, the mesonic contribution is smaller in lighter
nuclei. The mesonic contribution is calculated with the structure
function of ref. \cite{40}. We have also calculated it with the older 
structure functions of ref. \cite{56,57}. We find that around $x = 0.2$,  the
results for $R (x)$ decrease in about $3 \%$ if one uses the pion structure
function of ref. \cite{56} and increase in about $5 \%$ if one uses the one
of ref. \cite{57}.
This should give us an idea of the uncertainties of this contribution.
In fig. 11, for $^{56}Fe$, we split the mesonic contribution into the
pion and $\rho$-meson ones.
We observe, that although the pionic contribution
is bigger, the one from the $\rho$-meson
cloud is also important, and both of them are positive
in all the range of $x$. Similarly, as obtained in other calculations 
\cite{2,3,6} the mesonic contribution vanishes around $x = 0.6$ and increases
as $x$ decreases.
We can see that thanks to the mesonic contribution the agreement with the data
becomes much better. The slope of $R(x)$ from $x = 0.15$ to $x = 0.6$
is not reproduced by the contribution of the nucleons alone, a feature which
is shared by the results of ref. \cite{52}. The mesonic contribution comes to
produce the right slope.

As we said, the qualitative features of the pionic contribution are similar
to those in \cite{2,3,6}. There are also some differences. In ref. \cite{2}
the pionic contribution is evaluated assuming different amounts of pion 
excess in the nucleus, but no evaluation of this excess is made. Furthermore one
should recall our warnings in section 4 not to use the excess number in the
evaluation. In ref. \cite{3} an actual evaluation is done of the pionic 
contribution. Even if the formalisms here and there might look quite different,
they are actually quite similar and one can see that
$Im D (q) \equiv | D (q)^2| \; Im \Pi_{\pi}$ of eq. (48) appears in ref. \cite{3}
as $Im \Pi_{\pi} |D_0 (q)|^2$, with $|D (q)|^2$ changed to $|D_0 (q)|^2$, 
which is not very problematic when the pions are off shell. Furthermore, the
$\Delta h$ contribution is obtained from an extrapolation of the pionic atom
data, which is more problematic when one goes to the off shell situations
which one finds here. Furthermore, this $\Delta h$ is taken real
in \cite{3}, while here the $\Delta h$ 
Lindhard function is explicitly evaluated as a function of $q^0, q$ by
keeping the $\Delta$ width. An accurate evaluation of the imaginary parts of the 
diagrams is necessary in order to fulfill the sum rule of eq. (69).
On the other hand what one evaluates in ref. \cite{3} is the fractional
increase of $R(x)$ with respect to the one in the Sullivan process
(deep inelastic scattering with the pion cloud of a free nucleon) \cite{58,59}
and not the absolute value of the contribution of the pionic cloud.

In ref. \cite{6} the absolute contribution of the pion cloud is obtained by
subtracting the `` free '' parts  contained in the response function of a free
nucleon, as done here. The formalism is similar to ours but some approximations
are done which are improved here. For instance the $\Delta h$ contribution is
again taken real and other approximations are done to relate some magnitudes
to the pion excess number (recall our warnings about this).

The input which we have used for the meson nucleus interaction, 
$V_L (p), V_T (p)$, etc,
is the one of section 6. It has the virtue of having been tested in a large
variety of reactions and we have not changed it here. This gives us much
confidence about the strength of the mesonic contribution obtained here. The 
fact that it fills up the part of $R (x)$ missed by the nucleonic contribution
is certainly a welcome feature which reinforces our confidence on this mesonic
model.

The agreement of the results that we obtain with the experimental data
can be considered rather good by comparison with results
obtained with other theoretical approaches. The trend of the data is well
reproduced and the remaining discrepancies are not incompatible with the
intrinsic theoretical uncertainties of our model, particularly of the mesonic
contribution and more specially of the $\rho$-meson cloud which is somewhat
tied to the form factors and the nuclear correlation function, the fact that
we divide by $F_{2N} (x)$ instead of $F_{2D} (x)/2$ ,etc.

Our results for $R(x)$ in $^{40}Ca$ and $^{56}Fe$ look already very similar
and we have checked that $R(x)$ does not change much for heavier nuclei.
Obviously one should take into account that for heavier nuclei $N \neq Z$
and our approach with symmetric nuclear matter should be less accurate.
Actual calculations keeping different neutron and proton densities
lead to a slight decrease of the minimum of $R(x)$ as $Z$ increases \cite{53}.

\section{Conclusions}
We have evaluated the nucleonic and mesonic contributions to the ratio 
$R (x)$
of the $EMC$ effect, with particular emphasis on an accurate treatment of
effects shown in the past to be relevant, like binding effects,
Fermi motion and a dynamical (non static) treatment of the nucleons and the
mesons in the medium.

In order to avoid having to take some prescription on how to include relativistic
effects in the approach, which has led to many discussions in the past, we
started with a relativistic approach from the beginning and have made a
covariant treatment which allows us to write all magnitudes in terms of
nucleon and meson propagators in the medium. The approach was made practical
by evaluating the relevant magnitudes in an infinite medium and calculating
the structure functions in finite nuclei by means of the local density 
approximation. This procedure is fine for $x > 0.1$ but certainly breaks
down for $x < 0.1$ where there is nuclear shadowing. 

We could see that the use of the spectral functions to construct the nucleon
selfenergy was relevant in reducing somewhat the ratio $R (x)$ with respect
to a static picture of the nucleus, like a shell model, or in our case an
uncorrelated Fermi sea. This reconfirmed qualitatively earlier findings in 
the same direction.

Although our results for the nucleonic contribution differ somewhat from other
results in the literature, we share their conclusions that the nucleonic 
contribution alone does not explain  the data, particularly the slope from
$x = 0$ to $x = 0.6$.

On the other hand, we evaluated the contribution from the pion
and $\rho$-meson clouds rather
accurately. Recent work done before on the contribution of the pion cloud to
the $K^{+}$ nucleus selfenergy had taught us some important lessons, 
particularly the importance of using an input which satisfies a sum rule,
not trivial to satisfy unless the analytical properties of the pion selfenergy
are strictly fulfilled, and the need to avoid any relationship to the 
``pion excess number''. In addition, experience gained in dealing with
reactions which involve real and virtual pions allowed us to use
information on the pion nucleus selfenergy which is realistic enough and
has been tested in many such reactions. Hence, we consider the present
calculation of the pionic effects as an improvement over work done in the
past and we think these results are rather reliable. However, there are
still small uncertainties in the pionic contribution stemming from different
results for the pion structure function obtained in different analyses of
the Drell-Yan process.

The strength of the pionic effects is moderate.
So is the one from the $\rho$-meson cloud, but, when they are added to the 
nucleonic contribution one obtains a good description of the data in the
region of $0.1 < x < 0.7$.

In summary we could conclude that the main features of the $EMC$ effect
can be described in terms of conventional degrees of freedom, nucleons and
mesons. It does not exclude explanations in terms of more elementary degrees
of freedom like quarks and gluons. It is simply a question of which degrees
of freedom are more economical and transparent, as stressed by Jaffe in ref.
\cite{14}. The fact that we could deal with these degrees of freedom with
a certain accuracy, and establish the relationship of the effects found to familiar concepts
in conventional nuclear physics, makes these degrees of freedom rather 
appropriate to look at the $EMC$ and related effects.

\vspace{3cm}

Acknowledgements \\
We would like to acknowledge useful discussions with F. Gross, S. Liuti,
C. Garc\'{\i}a-Recio, A. Polls and V. Vento.

\vspace{0.5cm}

This work has been partially supported by CICYT contract number AEN 93-1205.
One of us E. M. wishes to acknowledge a fellowship from the Minis-\\
terio de Educaci\'on y Ciencia.

\newpage

\begin{center}
\underline{Figures captions}.
\end{center}

\vspace{0.5cm}

$\bullet$ Figure 1: Selfenergy diagrams of the nucleon.\\

$\bullet$ Figure 2: Electromagnetic form factors for the cases

a) free nucleon, b) Fermi sea with B baryons.\\

$\bullet$ Figure 3: (a) Feynman diagram for deep inelastic electron-nucleon 
scattering and (b), electron selfenergy diagram associated.\\

$\bullet$ Figure 4: Electron selfenergy diagram accounting for electron-pion
deep inelastic scattering.\\

$\bullet$ Figure 5: Diagrams of the electron selfenergy including $1 ph, 
1 \Delta h, 1 ph 1 \Delta h$, etc.. \\

$\bullet$ Figure 6: Two diagrams (a) direct and (b) crossed, which contribute
to Compton $\gamma \pi$ scattering.\\

$\bullet$ Figure 7: Results of $R_N (x)$ for $^{56} Fe$. Solid line: 
using the spectral function; dashed line: using the uncorrelated Fermi sea. 
Experimental points from ref. \cite{60} (solids dots), ref. \cite{61} 
(open squares).\\

$\bullet$ Figure 8: Results of $R(x)$ for $^{6} Li$. Solid lines: whole
calculation including the nucleons and the mesons; dashed line: contribution
of the nucleons.
Experimental points from ref. \cite{62} (solid dots). Density for $^{6} Li$
from ref. \cite{27}. 

$\bullet$ Figure 9: Same as fig. 8 for $^{12} C$.
Experimental points from ref. \cite{62} (solid dots), ref. \cite{60} (open
squares).\\

$\bullet$ Figure 10: Same as fig. 8 for $^{40} Ca$. 
Experimental points from ref. \cite{63} (solid dots), ref. \cite{60} (open
squares). \\

$\bullet$ Figure 11: Results for $R(x)$ for $^{56}Fe$. Solid line: 
whole calculation including the nucleons and the mesons; dashed line:
contribution of the nucleons; dot-dashed line: contribution of nucleons 
plus pions.
Experimental points from ref. \cite{60} (solid dots), ref. \cite{61} (open
squares).

\newpage

\begin{center}
Table I
\end{center}

\vspace{0.5cm}

\begin{tabular}{llllc}
 & & $<$T$>$ [MeV] & $<|$E - M$|>$ [MeV] & $|\varepsilon_A|$ [MeV] \\
\hline \\

$^{6}Li$ & UFS & 9.3 & 22.1 & 5.2 \\
  & SF & 18.8 & 33.8 &  5.2 \\

  &  &  &  &  \\

$^{12}C$ & UFS & 13.7 $\;$ (17.0)  & 31.0 $\;$ (23.0)  & 8.0 \\
  & SF & 31.7 $\;$ (37.0)  & 50.4 $\;$ (49.0)  & 7.8 \\ 

  &  &  &  &  \\

$^{40}Ca$ & UFS & 16.0 $\;$ (16.5)  & 33.5 $\;$ (26.6)  & 8.5 \\
  & SF & 40.4 $\;$ (36.0)  & 58.6 $\;$ (52.1)  & 8.6 \\

  &  &  &  &  \\

$^{56}Fe$ & UFS & 16.1 $\;$ (17.0)  & 34.1 $\;$ (25.0)  & 8.9 \\
  & SF & 40.4 $\;$ (33.0)  & 58.6 $\;$ (49.8)  & 8.7 \\

\end{tabular}

\vspace{2cm}

UFS: uncorrelated Fermi sea. SF: spectral function.
The numbers in brackets correspond to those obtained in ref. \cite{52} 
respectively for the Hartree-Fock and spectral functions (the latter called
there SRC, from short range correlations).

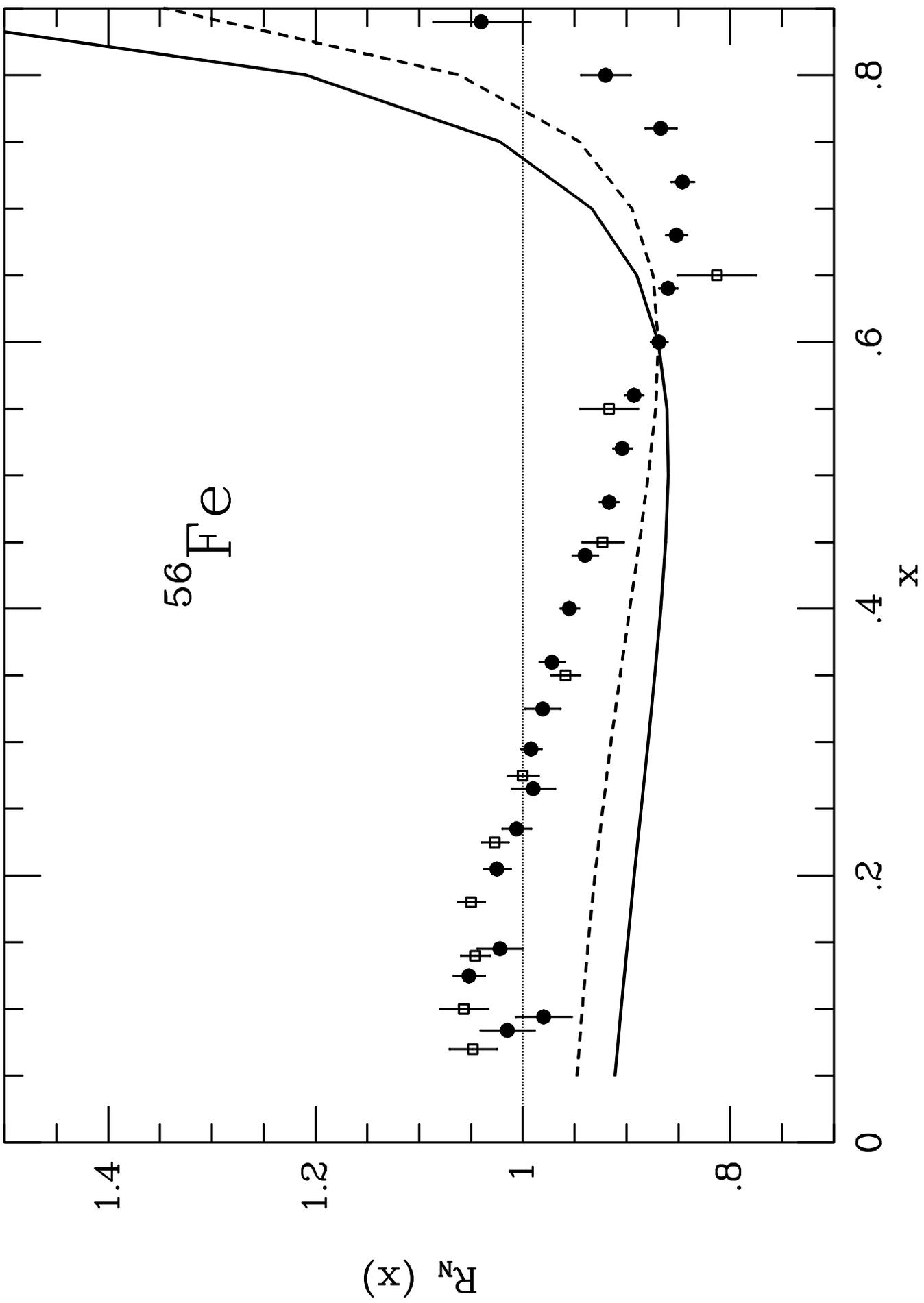
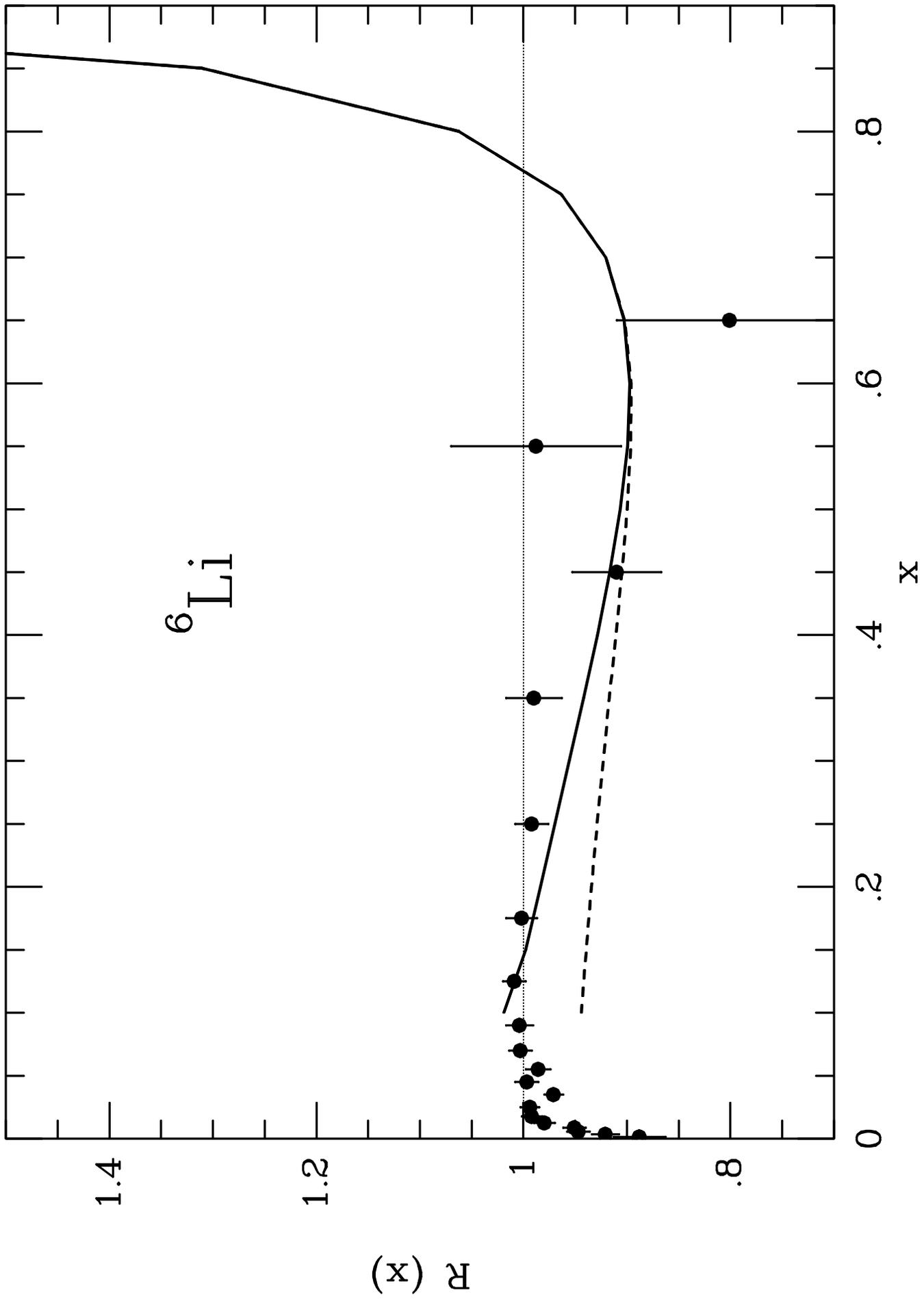
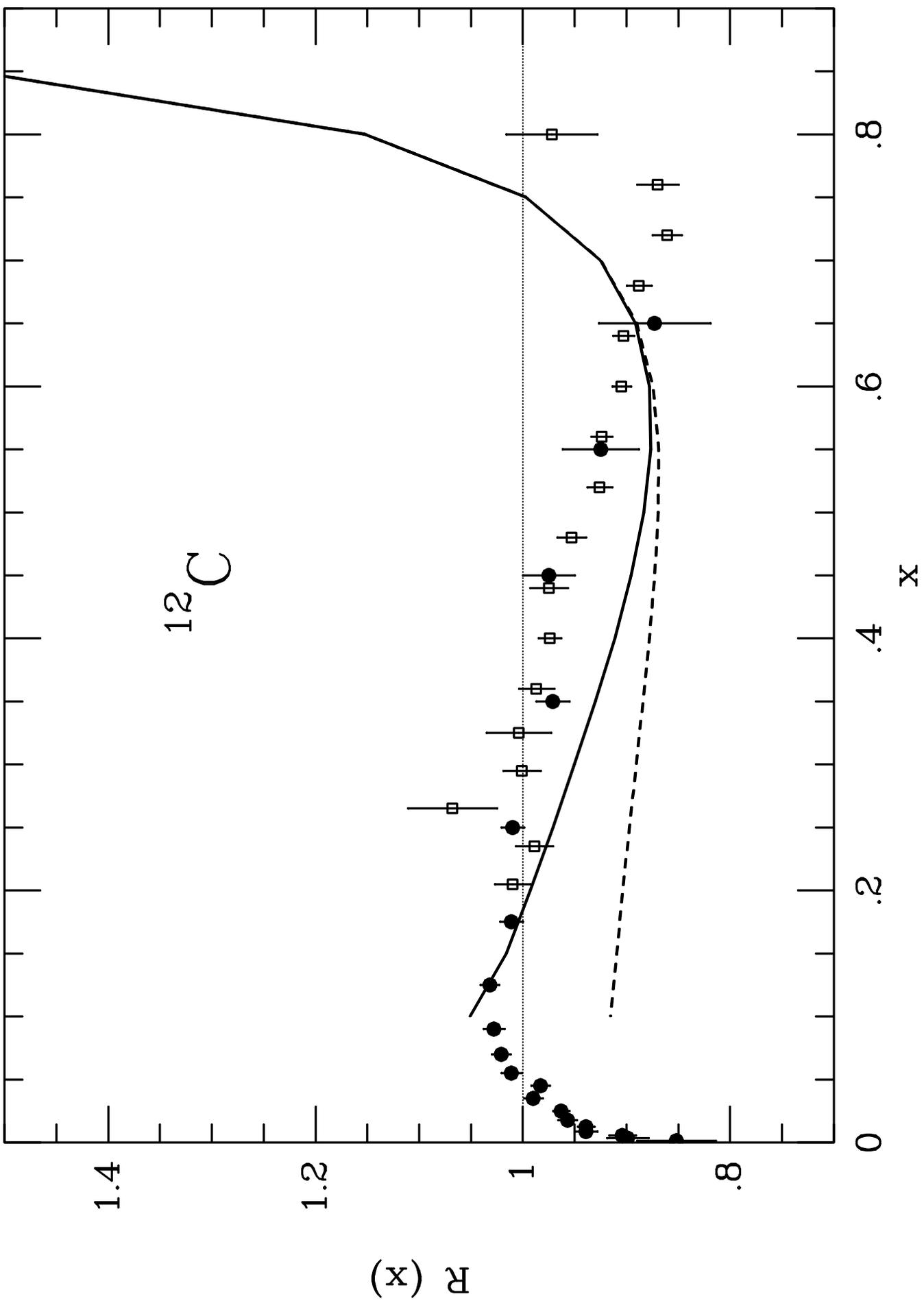
\end{document}